\documentclass[11pt,dvips]{article}
\usepackage{epsfig,times} 
\usepackage{picinpar}
\usepackage{wrapfig}
\usepackage{floatflt}
\makeatletter
\renewcommand{\section}{\@startsection{section}{1}{0in}
	{0.4\baselineskip}{0.1\baselineskip}{\Large\bf}}
\renewcommand{\subsection}{\@startsection{subsection}{2}{0in}
	{0.25\baselineskip}{-\baselineskip}{\large\bf}}
\renewcommand{\subsubsection}{\@startsection{subsubsection}{3}{0in}
	{0.1\baselineskip}{-\baselineskip}{\normalsize\bf}}
\makeatother
\begin{document}

%
\thispagestyle{myheadings}
%
\markright{HE.1.3.05}
\begin{center}
%
{\LARGE \bf Extensive air showers rounded up
by {\sc aires} + {\sc sibyll}/{\sc qgsjet}}
\end{center}

\begin{center}
%
%
{\bf L. A. Anchordoqui, M. T. Dova \& S. J. Sciutto}\\
{\it Departamento de F\'{\i}sica, Universidad Nacional de La
Plata \\
C.C. 67, (1900) La Plata \\ Argentina}
\end{center}
\begin{center}
{\bf  Abstract}
\end{center}
An analysis of the differences introduced by the hadronic
interaction event generators during the development of the
showers is presented. We have generated proton and nuclei induced
air showers with energies up to $10^{20.5}$ eV, ``herded up'' by 
the code {\sc aires} + {\sc sibyll}/{\sc qgsjet}. The most relevant 
observables are taken into account for the comparison.\\
\vspace{-0.5ex}
%
%
%


\vspace{1ex}

%
%

It is well known that extensive air shower (EAS) event
generators rely strongly on hadronic interaction models.
Mainly, because the first generation processes have
c.m. energies greatly exceeding those attained at man-made
accelerators, and thus, theoretical guidelines need to be used to
describe particle production. There are two codes with
algorithms tailored for efficient operation to the highest cosmic ray
energies. One of them was christened  {\sc sibyll} by  Fletcher 
et al. (1994). Its details of nucleus-nucleus interaction
were previously discussed by Engel et al. (1991, 1992). The other, 
{\sc qgsjet}, was performed by Kaidalov (1982), Kaidalov \&
Ter-Martirosyan (1982, 1984) and Kalmykov, Ostapchenko \& Pavlov (1997). 
See also, (Kaidalov, Ter-Martirosyan \&
Shabel'skii, 1986) for details of hadron-nucleus interaction,
and (Kalmykov \& Ostapchenko, 1993) for those of nucleus-nucleus
interaction. 

Recently, we have examined the sensitivity of
free parameters of these programs (which have been derived from
available collider data) when the algorithms are extrapolated several orders 
of magnitude (Anchordoqui et al., 1999); hereafter it will be referred as
paper I.
In particular, we have analyzed differences in the distribution of
depths of shower maximum, and the evolution of lateral and energy
distributions of showers induced by protons of $10^{20.5}$ eV. 
In this 
contribution we shall extend this analysis with results obtained from 
EAS initiated by heavy nuclei.

The nucleus-nucleus interaction is usually described using the
wounded nucleon picture in a Glauber multiple scattering framework 
(see,e.g., Bialas, Bleszynski \& Czyz, 1976; Pajares \& Ramallo, 1985). 
We shall adopt here the so-called ``semisuperposition''
model which retains the original idea that a shower induced by a nucleus 
may be modeled by the superposition of $A$ nucleon showers, but uses a 
realistic distribution of the positions of their first interaction.
To put into evidence as much as possible the differences between the
intrinsic mechanism of {\sc sibyll} and {\sc qgsjet} we have always
used the same code to simulate the fragmentation of the projectile, 
namely, the Hillas Fragmentation algorithm (Hillas, 1979, updated in 1981).
Differences introduced by primary fragmentation codes will be discussed 
elsewhere (work in progress).


Giant air showers induced by protons and nuclei with energies up
to $10^{20.5}$ eV were generated
with the code {\sc aires} (Sciutto, 1999), a realistic air shower 
simulation system which includes
electromagnetic interactions algorithms and links to the
mentioned {\sc sibyll} and {\sc qgsjet} programs.
Most of the electromagnetic algorithms are based on
the well known {\sc mocca} simulation program by Hillas (1997).

As in the paper I, in all the cases we have used the {\sc
aires} cross section, and all shower particles with energies
above the following thresholds were
tracked: 500 keV for gammas, 700 keV for electrons and positrons, 1
MeV for muons, 1.5 MeV for mesons and 80 MeV for nucleons and nuclei.
The particles were injected at the top of the atmosphere (100
km.a.s.l) and the ground level was located at sea level. All hadronic 
collisions with projectile energies below
200 GeV are processed with the Hillas Splitting algorithm (Hillas,
1979, 1981), 
and the external collision package is invoked for all those collisions with 
energies above the mentioned threshold.

Although $^{56}$Fe is certainly the best candidate for bottom up
acceleration mechanisms, at extremely high energies (around 200 EeV) 
the cosmic background radiation makes the universe opaque to the
propagation of iron nuclei, yielding severe constraints on the distance 
to the sources, as well as on the primary chemical
composition. Based on our previous analysis (Anchordoqui et al., 1998)
we have evaluated the photodisintegration of iron nuclei
($E \geq 5 \times 10^{19}$ eV) after a propagation distance
of 3 Mpc. The results, listed in Table I, were be taken into
account when computing the shower maximum energy dependence.

\begin{figure}[tb]
\label{fpks}
\centering
\leavevmode\epsfysize=6.8cm \epsfbox{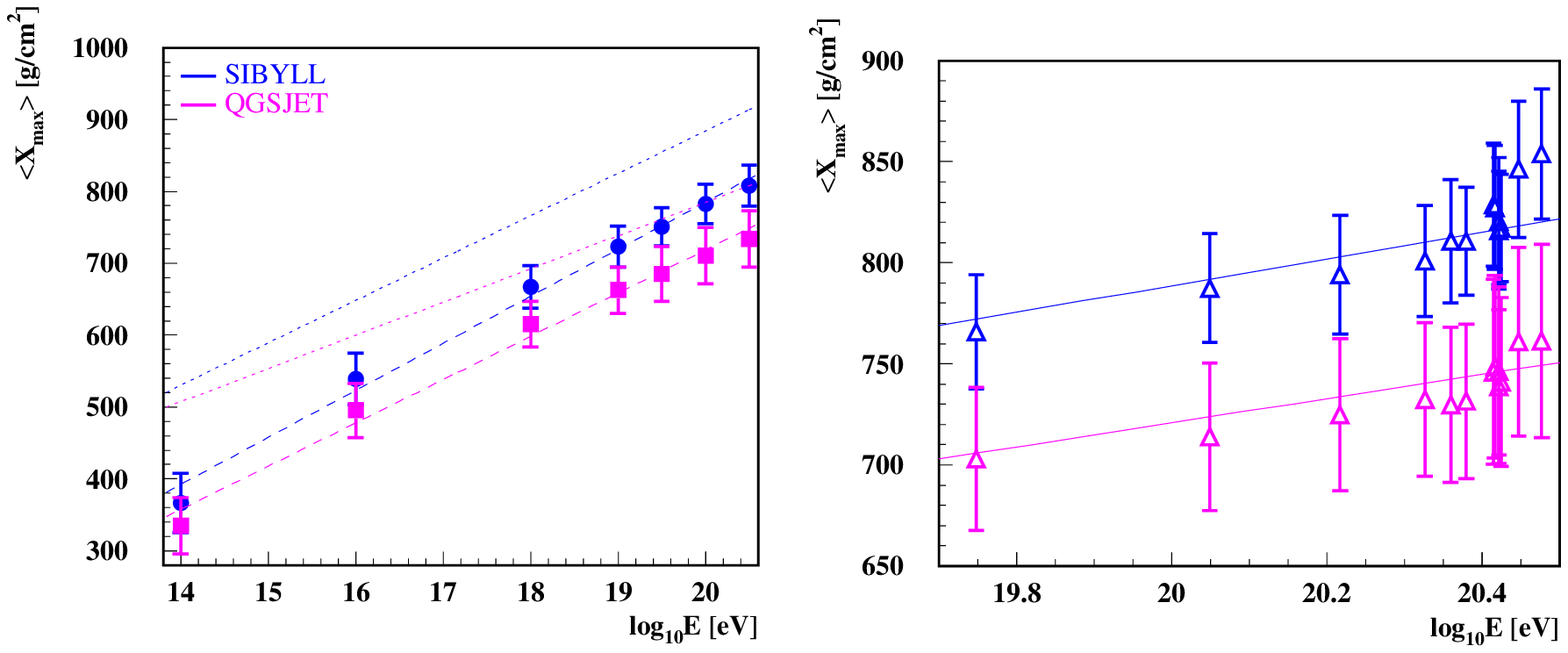}\\
\caption{Average slant depth of maximum. Left hand side (a) -- iron nuclei,
right hand side (b) -- heavy nuclei. (See the text).}
\end{figure}

In Fig. 1a we present the simulation results for the average slant
depth of maximum, $<X_{\rm max}>$, for iron nuclei induced showers. 
The error bars indicate the standard fluctuations 
(the rms fluctuations of the means are always smaller than the symbols).
It is evident that {\sc aires}+{\sc sibyll} showers present higher 
values for the 
depth of maximum, the differences increasing with rising energy. This
is consistent with the fact that in the first interaction {\sc sibyll} 
produces less secondaries than {\sc qgsjet}, yielding a delay in the
electromagnetic shower development which is strongly correlated with
decays of neutral pions. Besides, as it is 
expected, at the same total energy an air shower from a heavy
nucleus develops faster than a shower initiated by a proton (the
reader is referred to Fig. 7 of paper I). 
We have also computed estimations for the elongation rate, $d<X_{\rm
max}>/d \log_{10}E$, by means of linear fits to the data presented in
Fig. 1a. The slopes of the fitted lines are 65.47 g/cm$^2$ per decade and
60.23 per decade for {\sc aires}+{\sc sibyll} and {\sc aires}+{\sc qgsjet} 
respectively. Additionally, the dotted lines stand for the fits to
$<X_{\rm max}>$ for proton induced showers. In this case
the slopes are: 58.98 g/cm$^2$ for {\sc aires}+{\sc sibyll} and 
46.28 g/cm$^2$ for {\sc aires}+{\sc qgsjet}.
Around $10^{20}$ eV the primary chemical composition remains hidden by the
hadronic interaction model. Notice that at such a huge energy, proton showers
simulated with {\sc aires}+{\sc qgsjet} yield similar $<X_{\rm max}>$ that
the corresponding simulation of iron showers with {\sc aires}+{\sc sibyll}. 
Figure 1b, shows the results obtained after simulating heavy nuclei
(those listed in table I)
showers, together with the fits for the elongation rate of $^{56}$Fe 
induced showers. In this ``quite realistic'' scenario, the determination
of the chemical composition is even more dramatic.

In Fig. 2 we repeat the comparisons already performed 
in paper I. The behavior of the evolution of
electron-positron (first row of Fig. 2),
muon, and gamma lateral distributions, do not
show essential differences with respect to our previous analysis in paper I.
As in the case of protons,
despite the fact that
the high altitude lateral distributions
vary considerably with the hadronic interaction model,  the differences
seems to ``thermalize'' as long as the shower front
gets closer to the ground level. The second row stands for the
different particles energy distributions at sea level.
Except for  slight divergences in the muon case, again, the differences in
energy distributions at sea level do not correspond with deviations at
higher levels.

Putting all together, we found that the differences
between {\sc aires}+{\sc sibyll} and {\sc aires}
+{\sc qgsjet} at the surface, cannot make realize the original
differences present in the first generation of particles.

We turn now to the comparison of the recorded data between different
primary nuclei. In Fig. 3  it is shown the muon lateral
distributions for $^{12}$C and $^{56}$Fe (again
black stands for {\sc qgsjet} and grey for {\sc sibyll}).
Notice that there are no significant differences between the lateral
distributions at sea level when changing the chemical composition.
Nonetheless, the different predictions in the ground muon lateral
distribution, induced by the hadronic interaction models are
still present. Specifically, at 1000 m from the shower core
the ratio between the number of muons produced by
{\sc aires} + {\sc sibyll}/{\sc qgsjet} is 0.60
in a $^{56}$Fe induced shower and 0.62 in the case of $^{12}$C. 
Concerning the
number of electrons and positrons (at the same distance from the core),
the ratio between {\sc aires}+{\sc sibyll } and {\sc aires}+{\sc qgsjet}
predictions is 1.01 for iron nuclei, and 1.16 for carbon. Thus,
comparing these results with the ones obtained in paper I we observe that
the differences between the models diminish. This could be easily understood
if we recall that the differences in single collision between the models
increase with rising energy (see Sec. II of paper I). Now it is
straightforward that the heavier the nuclei the lower the energy per nucleon
in the first generation of particles.

\begin{figure}
\centering
\leavevmode\epsfysize=14cm\epsfxsize=15.5cm\epsfbox{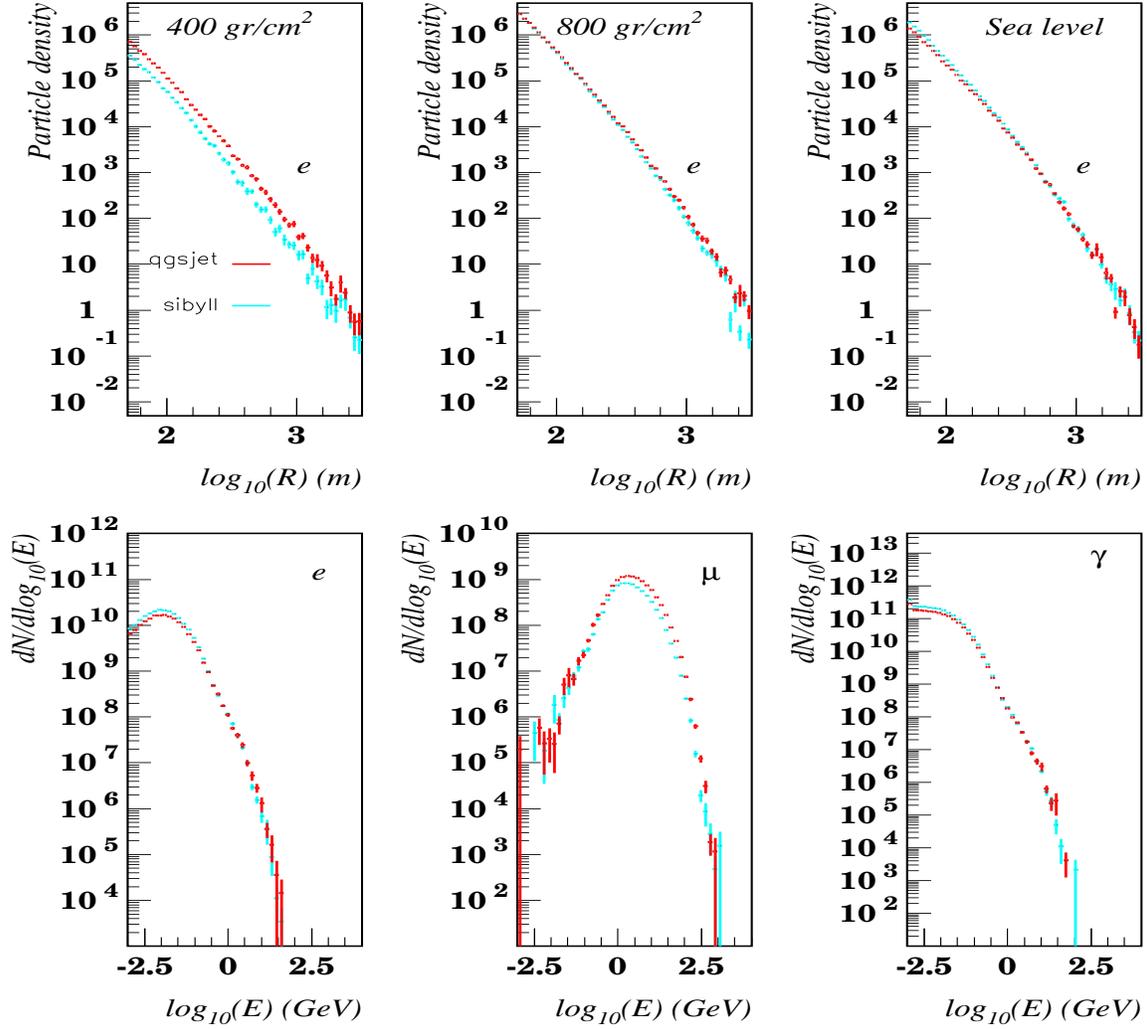}\\
\caption{Main features of iron nuclei shower development}
\end{figure}

\begin{figure}
\leavevmode\epsfysize=8.5cm\epsfxsize=17cm\epsfbox{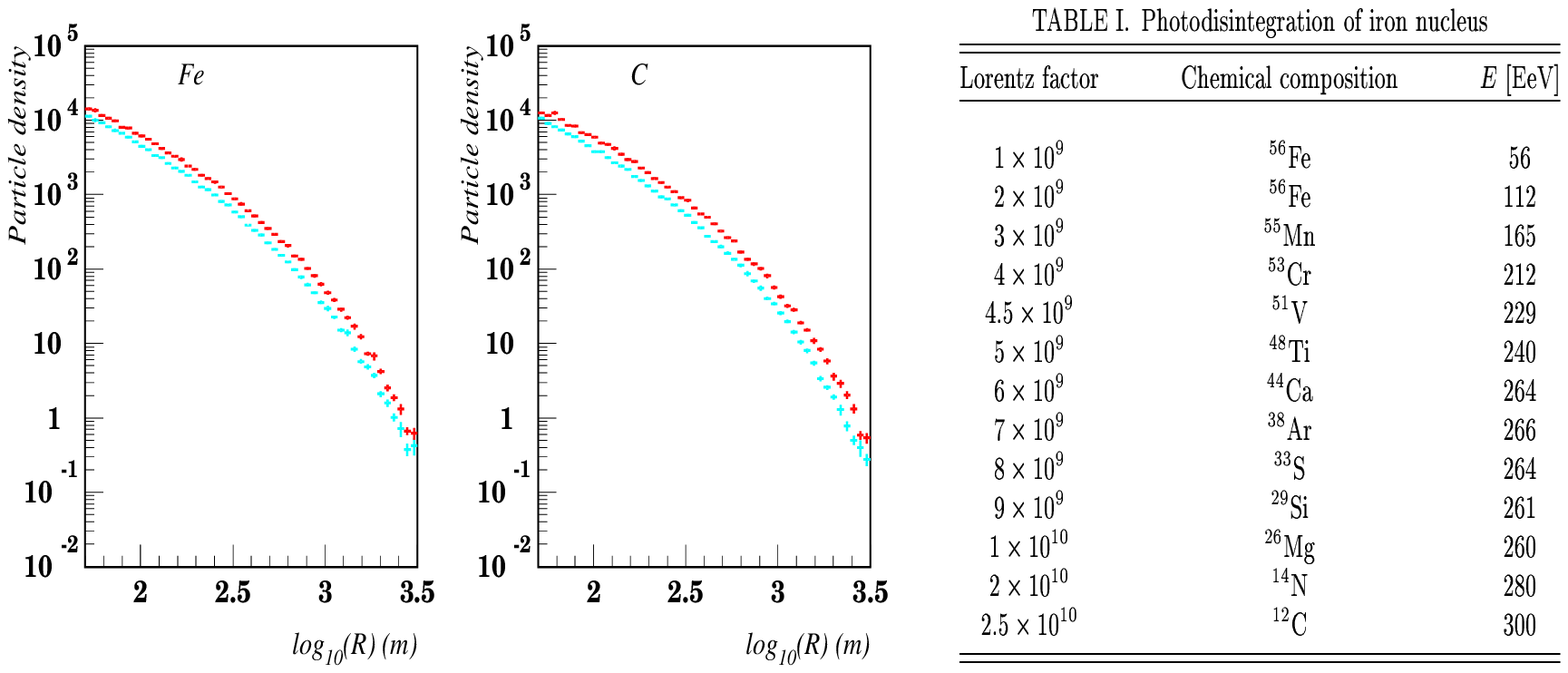}\\
\noindent Figure 3: Muon lateral distributions at sea level
\end{figure}
%
%
%
%
%
%
%

\vspace{1cm}

\vspace{1ex}
\begin{center}
{\Large\bf References}
\end{center}
%
Anchordoqui L.A. et al., 1998, Phys. Rev. D57, 7103\\
Anchordoqui L.A. et al., 1999, Phys. Rev. D59, 094003\\
Bialas A., Bleszynski M. \& Czyz W., 1976, Nucl. Phys. B111, 461\\
Engel J. et al., 1991, {\it Proc. of the 22$^{nd}$ International Cosmic Ray Conference}, Dublin, Ireland, Vol.4,p.1\\
Engel J. et al., 1992, Phys. Rev. D46, 5013\\
Fletcher R.S. et al., 1994, Phys. Rev. D 50, 5710\\
Hillas A.M., 1979, {\it Proc. of the 16$^{\rm th}$
International Cosmic Ray Conference}, Tokyo, Japan,  Vol.8,p.7\\
Hillas A.M., 1981, {\it Proc. of the 17$^{\rm th}$ International Cosmic Ray Conference},
Paris, France, Vol.8,p.183\\
Hillas A.M., 1997, Nucl. Phys. B (Proc. Suppl.) 52, 29\\ 
Kaidalov A.B., 1982, Phys. Lett B116, 459\\
Kaidalov A.B. \& Ter-Martirosyan K.A., 1982,
Phys. Lett. B117, 247\\
Kaidalov A.B. \& Ter-Martirosyan K.A., 1984,
Yad. Fiz. 39, 1545 [Sov. J. Nucl. Phys. 39, 979]\\
Kaidalov A.B., Ter-Martirosyan K.A \&  Shabel'skii Y.M., 1986,
Yad. Fiz. 43, 1282 [Sov. J. Nucl. Phys. 43, 822]\\
Kalmykov N.N.,  Ostapchenko S.S. \& Pavlov A.I., 1997,
Nucl. Phys. B (Proc. Supp.) B52, 17\\
Kalmykov N.N. \&  Ostapchenko S.S., 1993, Yad. Fiz. 56, 105
[Phys. At. Nucl. {\bf 56}, 346]\\ 
Pajares C. \&  Ramallo A.V., 1985, Phys. Rev. D31, 2800\\ 
Sciutto S.J., 1999, {\sc aires:} {\it A System for Air Shower
Simulations}, Auger technical note GAP-99-020\\

\vspace{1ex}

\noindent{\small This article will be published in {\it Proc. 26th
International Cosmic Ray Conference} (Salt Lake City, Utah, 1999).}

\end{document}